\documentclass[twocolumn,prl,aps,showpacs]{revtex4}

\usepackage{psfrag,graphicx}

\usepackage{dcolumn}
\usepackage{amsmath,amssymb}
\usepackage{bm}
\usepackage[dvips]{color}

\definecolor{mygrey}{gray}{0.35}
\definecolor{mygreen}{rgb}{0.85,1,0.9}
\definecolor{myzard}{cmyk}{0,0,0.05,0}
\definecolor{mywhite}{rgb}{1,1,1}
\definecolor{myred}{rgb}{1,0,0}

\def\R{{\mathbb{R}}}

 \def\ee{\mathord{\rm e}}
 
 \def\ii{\mathord{\rm i}}

\def\half{\textstyle\frac{1}{2}}

\def\vec#1{{\bf{#1}}} 

\def\bra#1{\langle#1|} \def\ket#1{|#1\rangle}

\begin{document}

\title[Short Title]{
Entanglement and Concurrence in the BCS State}

\author{M.A. Mart\'{\i}n-Delgado}
\affiliation{
Departamento de F\'{\i}sica Te\'orica I, Universidad Complutense,
28040. Madrid, Spain.
}

\begin{abstract}
An extension of the notion of concurrence introduced by Wooters
is used to quantify the entanglement of certain multipartite pure states,
namely, the BCS state of superconducting compounds. This leads to a
definition of the macrocanonical entanglement of pairing (MEP) for which we
compute an analytical formula in terms of two adimensional numbers,
the cut-off and gap numbers, which depend on measurable physical quantities.
We find that strongly coupled BCS elements like Pb and Nb have much larger
MEP values than more conventional BCS transition metal superconductors.
\end{abstract}

\maketitle

The study of entanglement of quantum states is now a central issue in
the modern theory of quantum information \cite{books}.
Entanglement is not seen just
as a sort of peculiar or curious property of multipartite quantum states, but more
importantly, it has become a resource to achieve novel tasks, such as
teleportation, quantum crytography
and other quantum transmission protocols surpassing
the capabilities exhibited by their classical counterparts. Moreover,
entanglement is also something of direct experimental importance since
it amounts to interaction among parties: just to mention a simple
instance of this,
the application of the CNOT gate to a factorizable state
$\ket{\Psi}_{\rm in}:=2^{-1/2}[\ket{0}-\ket{1}]\ket{1}$
produces one of the Bell states, namely,
$U_{\rm CNOT}\ket{\Psi}_{\rm in}=
2^{-1/2}[\ket{01}-\ket{10}]:=\ket{\Psi^-}$. In this process, the output
state is more difficult (and expensive) to realize than the initial state
$\ket{\Psi}_{\rm in}$ because of the CNOT gate that implements entanglement.
Therefore, to know wheather  a given purported state is entangled or not,
as well as how much entangled it is, is of great importance both theoretically
and experimentally.

In condensed matther theory, we are used to deal with quantum many-body states
(or multiqubit states)
in which  their strong quantum correlations are responsible for
novel properties or states of matter, like quantum liquids, or quantum
phase transitions \cite{jaitisi},\cite{strongly}.
Example of strongly correlated states abound in these areas, for example,
valence bond states are nothing but Bell states, and  properties
like entanglement swapping correspond to resonating valence bond (RVB)
state configurations.

In low dimensional systems, like quantum spin chains and ladders, it is
known that the effect of quantum fluctuations is stronger than in higher
dimensions: the factorizable Neel state is a good
starting point to describe the ground state of the antiferromagnetic
Heisenberg model in $D=2$ or more dimensions, but it is unsuitable
for $D=1$ where the Bethe ansatz solution is a complicated superposition
(entangled) of single particle states. Thus, it is natural to ask whether
the new ideas about qualifying and quantifying entanglement that have
emerged in the field of quantum information can be helpful to describe
the complicated patterns of behaviour exhibited by strongly correlated
systems in condensed matter.

Recently, the entanglement properties of the one dimensional XY model in
a transverse magnetic field
have been analyzed in the vicinity of a
quantum phase transition  by Osterloh et al. \cite{osterloh}
and Osborne and Nielsen \cite{bertin}.
In this model, entanglement shows
scaling behaviour near the transition point and remains
short ranged \cite{osterloh}. The quantification of entanglement is made
with the entanglement measure known as {\em entanglement of formation}
$E(\rho)$,
introduced by Bennett et al. \cite{bennett1} to describe the resources
needed to create a given entangled bipartite state,
either pure $\ket{\psi}$ or mixed $\rho$ \cite{bennett2}.

Generally, it is difficult to find closed mathematical expressions of $E(\rho)$
solely in terms of $\rho$, but for the special case of mixed states of
bipartite qubit systems Wooters \cite{wooters}, \cite{hill-wooters} found
one such a formula. This formula makes use of what Wooters calls \cite{wooters}
a {\em spin flip transformation}, defined as
\begin{equation}
\ket{\widetilde{\psi}}:=\sigma_y \ket{\psi^{\ast}}
\label{0a}
\end{equation}
where $\textstyle \sigma_y:=\begin{pmatrix}
0 & -\ii \\
\ii & 0
\end{pmatrix}$ is the Pauli matrix in the computational basis
$\{\ket{0}:=\ket{\kern -4.5pt\uparrow},
\ket{1}:=\ket{\kern -4.5pt\downarrow}\}$.
For a general state $\rho$ of two qubits, the spin-flipped state
is
$\tilde{\rho}:=(\sigma_y\otimes\sigma_y)\rho^{\ast}(\sigma_y\otimes\sigma_y)$.
When the state is pure $\rho=\ket{\psi}\bra{\psi}$, the entanglement of
formation  can be written as $E(\rho)={\cal E}(C(\psi))$, where the
concurrence $C$ is defined as
\begin{equation}
C(\psi):=|\langle\psi\ket{\widetilde{\psi}}|
\label{0b}
\end{equation}
and ${\cal E}(C):=h(\half[1+\sqrt{1-C^2}])$,
$h(x):=-x\log_2x-(1-x)\log_2(1-x)$, is a monotonically increasing function
of $C$ that ranges from 0 to 1 as the concurrence goes from 0 to 1.

In fact, Wooters proposes to use concurrence (\ref{0b}) as an entanglement
measure in its own right. Here, we shall adhere to this proposal by studying
the concurrence of a physically realizable state such as the BCS state
\cite{bcs}.
There are very few cases where we have a solution to a quantum
many-body problem in the form of an explicit wave function. The BCS
theory of standard superconductors provides us with one of these examples.
Specifically, the general quantum state representing a superconductor carrying
a supercurrent at T=0 temperature is

\begin{equation}
\ket{\rm BCS}_{\theta} :=
\prod_{\vec{k}}
(u_k + \ee^{\ii \theta} v_k
c^{\dagger}_{\vec{k}+\vec{Q},\uparrow}
c^{\dagger}_{\vec{-k}+\vec{Q},\downarrow}) \ket{0}
\label{1}
\end{equation}
where $\ket{0}$ denotes the zero-particle Fock state. In this state the
electrons are created in Cooper pairs for the occupied states with quantum
numbers $(\vec{k}+\vec{Q},\uparrow; \vec{-k}+\vec{Q},\downarrow)$.
All these pairs have the same momentum $2\hbar Q$. This pair momentum
represents the finite supercurrent and it is usually very small.
We shall concentrate in the static condensate of Cooper pairs with zero
supercurrent $Q=0$. The parameters $u_k$ and $v_k$ represent the probability
amplitudes of creating quasi-holes and quasi-electrons, respectively.
They satisfy the following properties:
\begin{equation}
\begin{split}
u_k^2 + v_k^2 = 1, \\
u_k, v_k \in \R; \ k:=|\vec{k}|\\
\end{split}
\label{2}
\end{equation}
The first condition comes from the normalization of the state
$\ket{\rm BCS}_{\theta}$, and they only depend on the modulus of $\vec{k}$.
The phase factor $\ee^{\ii \theta}$ is arbitrary, but is the same for all
Cooper pairs.
In the macrocanonical BCS state (\ref{1}), the number of Cooper pairs $N$
is not a well defined quantity. By series expansion,
the state can be thought of as an average
over an ensemble of states $\ket{N,Q}$ with a definite number $N$ and pair
momentum $2\hbar Q$:
$\ket{{\rm BCS}}_{\theta} := \sum_N \ee^{\ii N \theta} A_N \ket{N,Q}$,
$\sum_N A_N^2 = 1$.

The advantage of having the explicit form of the ground state
wave function is that we can compute any quantity needed for an entanglement
measure. In particular, it is possible to compute the reduced density matrix
$\rho(\vec{k}_i,\vec{k}_j)$ by tracing out over all Cooper pairs with momenta
$\vec{k}\neq \vec{k}_i,\vec{k}_j$. This is a bipartite density matrix
for which explicit formulas for the entanglement of formation
 also exist in terms of the concurrence \cite{wooters}.
However, we notice that in the case of the macrocanonical BCS state, the
matrix $\rho(\vec{k}_i,\vec{k}_j)$  corresponds precisely
to the state formed by the product of two Cooper pairs with momenta
$(\vec{k}_i,\vec{k}_j)$, namely,
$[u_{k_i}\ket{0}+v_{k_i}\ket{1}]\otimes[u_{k_i}\ket{0}+v_{k_j}\ket{1}]$,
where here $\ket{0},\ket{1}$ denote states with zero and one Cooper pair,
respectively. Then, the reduced density matrix leads to the same original
problem we are dealing with, but with only two pairs.

Despite of this difficulty, is it still possible to use the
concurrence to devise an entanglement probe for the
macrocanonical BCS state? In principle it looks difficult since
the BCS state is a many-body (multiqubit) state and it is known
that concurrence fails to capture entanglement properties of
multiqubit states. For example, it is known that any qutrit state
can be entangled in two different ways \cite{durr}: either as a
$\ket{\rm GHZ}:=2^{-1/2}(\ket{000}+\ket{111})$ state or as a Werner
$\ket{\rm W}:=3^{-1/2}(\ket{001}+\ket{010}+\ket{100})$ state.
Both of them yield zero concurrence since
$\ket{\widetilde{\rm GHZ}}=\ii2^{-1/2}(\ket{000}-\ket{111})$ and
$\ket{\widetilde{\rm W}}=\ii3^{-1/2}(\ket{110}+\ket{101}+\ket{011})$.
Thus, concurrence  does not detect the existence of entanglement 
for qutrits.

However, here we show that relying on physical grounds and
motivated by the physical meaning of the concurrence, it is
possible to give an entanglement measure for the BCS state based
on the notion of concurrence. We do this in two steps.

In what follows, we shall study a many-body state $\ket{\Psi}$
which is a BCS ground state with $\theta=0$:
\begin{equation}
\ket{\Psi}:=\ket{\rm BCS}_{0} =
\prod_{\vec{k}}
(u_k + v_k
c^{\dagger}_{\vec{k},\uparrow}
c^{\dagger}_{\vec{-k},\downarrow}) \ket{0}
\label{3}
\end{equation}
This BCS state is the solution of minimum energy to a reduced
Hamiltonian called pairing Hamiltonian \cite{schrieffer}:

\begin{equation}
H_{\rm red}:=
\sum_{\vec{k}} 2\epsilon_k b^{\dagger}_{\vec{k}} b_{\vec{k}} -
\sum_{k\neq k'} V_{\vec{k},\vec{k'}} b^{\dagger}_{\vec{k}'} b_{\vec{k}}
\label{4}
\end{equation}
where $b^{\dagger}_{\vec{k}}:=
c^\dagger_{\vec{k},\uparrow} c^\dagger_{-\vec{k},\downarrow}$,
$b_{\vec{k}}:=c_{-\vec{k},\downarrow} c_{\vec{k},\uparrow}$
are operators creating and annhilating Cooper pairs, respectively.
The solution to this variational problem yields the following expressions
for the probability amplitudes
\begin{equation}
\begin{split}
u_k^2 &= \half (1 + \frac{\epsilon_k}{E_k})\\
v_k^2 &= \half (1 - \frac{\epsilon_k}{E_k})\\
E_k &= \sqrt{\epsilon_k^2 + \Delta_k^2}
\end{split}
\label{5}
\end{equation}
where $E_k$ is the energy of the quasi-particles (excitations),
and $\Delta_k$ is called the gap function, which is determined by
the self-consistent solution of the  gap equation
$\Delta_k = -\sum_{\vec{k}'} \frac{\Delta_{k'}}{2E_{k'}} V_{\vec{k},\vec{k'}}$.
This solution represents a BCS superconductor or SC state.

In the first step, we extend the notion of concurrence to many-body
states based on the physical interpretation of concurrence introduced
by Wooters \cite{wooters}. Namely, for a spin $\half$ particle the
spin-flip operation (\ref{0a}) is the time-reversal operation. For two-qubit
states, we can argue that concurrence can serve as an entanglement measurement
directly form this definition, without resorting to its connection to
${\cal E}(C)$. The rationale goes as follows:
as $\ket{\widetilde{\psi}}$ is obtained from $\ket{\psi}$ by time inversion,
we intuitively expect that if $\ket{\psi}$ is very much entangled, then
$\ket{\widetilde{\psi}}$ will be very similar to $\ket{\psi}$ thereby
$C(\psi)\sim 1$. On the contrary, if $\ket{\psi}$ is factorized into
two states then
$\ket{\widetilde{\psi}}$ will be very different from  $\ket{\psi}$
and then $C(\psi)\sim 0$. When the given state is very entangled, the
time-revesed state is very close to the original state and their overlap
is very large.

Thus, it is reasonable to extend the notion of spin-flip opertation by
the time-reversal operation.
Let us define the concurrence $C(\rm BCS)$
of the BCS state (\ref{3})  by means of the
overlapping with its time-reversed state, namely,
\begin{equation}
C({\rm BCS}) := | _0\!\bra{{\rm BCS}}\widetilde{{\rm BCS}}\rangle_0|
\label{7}
\end{equation}
with
\begin{equation}
|\widetilde{{\rm BCS}}\rangle_0 := U_T \ket{{\rm BCS}}
\label{8}
\end{equation}
The action of the time-reversal operator $U_T$ on position, momentum and
spin variables is \cite{gp}
\begin{equation}
\begin{split}
U_T \vec{r}_i U_T^\dagger & = \vec{r}_i \\
U_T \vec{k}_i U_T^\dagger & = -\vec{k}_i \\
U_T \vec{s}_i U_T^\dagger & = -\vec{s}_i
\end{split}
\label{9}
\end{equation}

In the second step, we realize that in the absence of a precise connection
between entanglement of formation and concurrence beyond two-qubit states
as the one provided by Wooters \cite{wooters}, \cite{hill-wooters},
we must define an entanglement measure with respect to a reference state
for which we know that it has zero entanglement.
The natural candidate for this
is the Fermi Sea state $\ket{\rm FS}$ defined as
\begin{equation}
\ket{\rm FS}:=\prod_{k\leq k_{\rm F}}
c^{\dagger}_{\vec{k},\uparrow}c^{\dagger}_{\vec{k},\downarrow}
\ket{0}
\label{00}
\end{equation}
Using (\ref{9}), this state has maximum concurrence $C(\rm FS)=1$
despite being unentangled.
Thus, we choose as our definition of entanglement what we shall call
MEP {\em macrocanonical entanglement of pairing} $\Bbb{E}(\rm BCS)$ defined as
\begin{equation}
\Bbb{E}(\rm BCS):= \langle\rm FS\ket{\widetilde{FS}} -
\langle\rm BCS\ket{\widetilde{BCS}}
\label{9b}
\end{equation}
With this definition the Fermi Sea state has zero entanglement.
Now, we compute the MEP to check that it serves
as a record signalling and quantifying the onset of entanglement
in the macrocanonical BCS state.
Using the fact that $U_T\ket{0}=\ket{0}$ and (\ref{9}) along with the
canonical anticommutation relations, we find
\begin{equation}
|\widetilde{{\rm BCS}}\rangle_0 =
\prod_{\vec{k}}
(u_k - v_k
c^{\dagger}_{\vec{k},\uparrow}
c^{\dagger}_{\vec{-k},\downarrow}) \ket{0} =
|{\rm BCS}\rangle_{\pi}
\label{10}
\end{equation}
This means that under time-reversal, the state with $\theta=0$ is transformed
into another one with $\theta=\pi$.
Let us point out that all states $\ket{{\rm \widetilde{BCS}}}_{\theta}$
have the same energy with respect to the reduced Hamiltonian (\ref{4})
\cite{comment1}.

Now, we can compute the concurrence of the BCS state as the overlapping
between states (\ref{10}) and (\ref{3}), yielding
\begin{equation}
C({\rm BCS}) = \prod_k |(u_k^2 - v_k^2)|
\label{11}
\end{equation}
This suggests to introduce the notion of a partial concurrence $C_k({\rm BCS})$
for each Cooper pair of momentum $k$, and using the BCS solution (\ref{5})
to the gap equation we readly find
\begin{equation}
C_{k}({\rm BCS}) :=  |(u_k^2 - v_k^2)| =
|\frac{\epsilon_k}{\sqrt{\epsilon_k^2+\Delta_k^2}}|
\label{12}
\end{equation}
\begin{figure}[t]
\psfrag{x}[Bc][Bc][1][0]{$\epsilon_k - \mu$}
\psfrag{y}[Bc][Bc][1][90]{$C_k({\rm BCS})$}
\psfrag{a}[Bc][Bc][.65][0]{$0.0$}
\psfrag{b}[Bc][Bc][.65][0]{$0.1$}
\psfrag{c}[Bc][Bc][.65][0]{$0.5$}
\psfrag{d}[Bc][Bc][.65][0]{$0.9$}
\psfrag{g}[Bc][Bc][.65][0]{$\Delta$}
\includegraphics[width=8 cm]{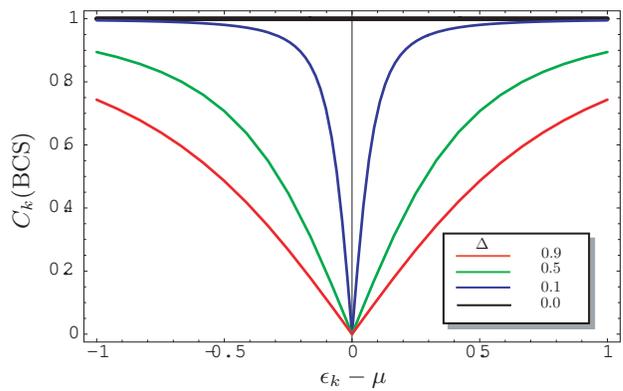}
\vspace{-15 pt}
\caption{Partial concurrences $C_k({\rm BCS})$ in terms of energy differences
w.r.t. the chemical potential $\mu$, for several values of the superconducting
gap $\Delta$.}
\label{fig1}
\end{figure}
In Fig.~\ref{fig1}
we plot $C_k({\rm BCS})$ as a function of the energy $\epsilon_k$
for several values of an homogeneous gap function $\Delta_k:=\Delta$.
It clearly shows that the partial concurrences are different from 1 in the
vicinity of the Fermi surface, and this region extends within an interval of
the order of $2\Delta$. This is precisely the region where Cooper pairs are
being formed. Thus, deviations of $C_k({\rm BCS})$ from 1 allows us to detect
the onset of correlations between pairs of particles. In fact, when the gap
vanishes the solution to $u_k,v_k$ in (\ref{5})
yields $C_k({\rm BCS})=1, \forall k$,
which agrees with the fact that it represents a normal metal with
an uncorrelated (factorizable) ground state (\ref{00}).

\begin{figure}[ht]
\psfrag{x}[Bc][Bc][1][1]{$n_1$}
\psfrag{y}[Bc][Bc][1][1]{$n_2$}
\psfrag{z}[Bc][Bc][1][1]{$\Bbb{E}({\rm BCS})$}
\includegraphics[width=8 cm,height=6 cm]{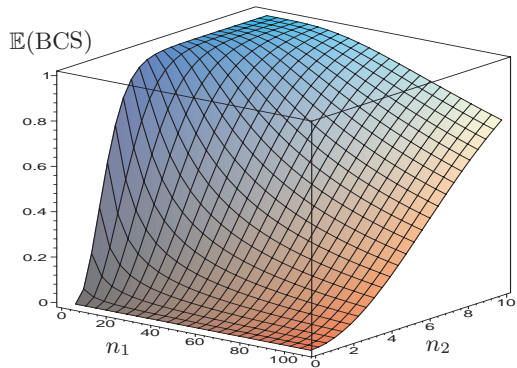}
\caption{Macrocanonical entanglement of pairing (MEP) $\Bbb{E}({\rm BCS})$
as a function of the cut-off number $n_1$ and gap number $n_2$ (\ref{13}).
The blue regions represent MEP high while red regions correspond to MEP low.}
\label{fig2}
\end{figure}

We can proceed even further and compute the total concurrence of the BCS state.
It can be computed exactly in the continuum limit
$\sum_{\vec{k}} \rightarrow \left(\frac{L}{2\pi}\right)^3\int d^3k$ with
the following result
\begin{equation}
\begin{split}
C({\rm BCS}) =&
\left[ 1 + \left(\frac{n_2}{n_1}\right)^2 \right]^{-\frac{n_1}{2}}
\ee^{-n_2\arctan\frac{n_1}{n_2}} \\
n_1  :=& N(\epsilon_{\rm F}) \hbar \omega_{\rm D}\\
n_2  :=& N(\epsilon_{\rm F}) \Delta
\end{split}
\label{13}
\end{equation}
where $N(\epsilon_{\rm F})$ is the density of states at the Fermi level
and $\omega_{\rm D}$ is the Debye frequency. For instance, for a parabolic
dispersion relation like $\epsilon_k:=\frac{\hbar k^2}{2m}$ we have
\begin{equation}
N(\epsilon_{\rm F}):=
\frac{L^3}{2\pi^2}\left[ k^2\frac{dk}{d\epsilon_k}
\right]_{\epsilon_k=\epsilon_{\rm F}} = L^3 \frac{mk_{\rm F}}{2\pi^2\hbar^2}
\label{14}
\end{equation}
Therefore, we find that the BCS concurrence (\ref{13})
depends on two adimensional
quantities with a physical origin that we call the {\em cut-off number} $n_1$
and the {\em gap number} $n_2$.
From (\ref{13}) we see that this concurrence is always
$\leq 1$, and the maximum value is attained when $n_2=0$ corresponding to
the absence of superconductivity (and thus, no correlated pairs).
In Fig.~\ref{fig2} we plot the macrocanonical entanglement of pairing MEP
(\ref{13}) to show how it depends on the numbers $n_1$ and $n_2$. We see
that for a fixed value of $n_1$ (and thus of the cut-off $\omega_{\rm D}$
and $N(\epsilon_{\rm F})$),
the entanglement increases with $n_2$ and in turn with the superconducting
gap $\Delta$ at $T=0$.

The nice feature about the analytical formula (\ref{13}) for the
entanglement is that it depends
through $n_1$ and $n_2$ on the experimentally accessible quantities
$N(\epsilon_{\rm F})$, $\omega_{\rm D}$ and $\Delta$. We propose to
use MEP in superconducting compounds as an indicator of the entanglement
or quantum correlations due to the presence of Cooper pairs.
It is interesting to start with the simplest cases, namely,
the superconductive elements. In Table~\ref{tableMEP} we have computed
the values of MEP from available experimental data
\cite{CRC}, \cite{vonsovsky}, \cite{roberts}
for transition superconducting metals and lead. We find relevant to
put the MEP values
together with the dimensionless electron-phonon coupling constant $\lambda$.
In the conventional BCS theory, we have $\lambda\ll 1$
and the effect of phonons is to provide a  cut-off $\hbar \omega_{\rm D}$
to the possible electron energies. However, there are materials like Pb and Nb
for which retardation effects of phonons are relevant for the pairing of
electrons, and they need to be treated dynamically within the
Eliashberg extension of the BCS theory \cite{scalapino}.
In this theory $\lambda$ is defined
as $\lambda:=2\int_0^{\infty}d\omega \alpha^2(\omega)N_{\rm ph}(\omega)/\omega$
where $N_{\rm ph}(\omega)$is the phonon density of states and
$\alpha^2(\omega)$ is the electron-phonon coupling strength. In the
weak coupling limit $\lambda\ll 1$, it reduces to the BCS coupling parameter
$\lambda \approx N(\epsilon_{\rm F})V_0$.

From this table a clear picture emerges: as $\lambda$ gets bigger, there is
a large increase in the value of MEP, specially for the case of Eliashberg
superconductors which have values of MEP about 3 orders of magnitude higher
than in more conventional superconductors.
For the group of elements Ru, Mo and Os that have similar values of $\lambda$,
their correspoding MEP values are also very close.
Thus, we conclude that strongly coupled BCS superconductors are characterized
by large MEP values. We find this reasonable since the effect of phonons is
to enhance the electronic correlations \cite{scalapino}. This enhancement
is also responsible of the higher $T_c$ values in Table~\ref{tableMEP}.
\begin{table}[t]
\begin{ruledtabular}
\begin{tabular}{cccc}
 {\rm Superconductive Elements} & $-\log_{10}(\Bbb{E}({\rm BCS}))$ &
$\lambda$ & $T_c$(K)\\
 \hline \hline
{\rm Transition SC Metals} & &
\\ \hline \hline
{\rm Hf} & 7.557 & 0.14 & 0.13\\
{\rm Ru} & 6.813 & 0.38 & 0.49 \\
{\rm Mo} & 6.381 & 0.41 & 0.92 \\
{\rm Os} &6.290 & 0.44 &  0.66\\
\hline \hline
{\rm Eliashberg SC} & & \\
\hline \hline
{\rm Nb} & 3.549 & 0.82 &  9.25\\
{\rm Pb} & 3.295 & 1.55 &  7.20\\
\end{tabular}
\end{ruledtabular}
\caption{A list of superconductive elements with their values
of macrocanonical entanglement of pairing $\Bbb{E}({\rm BCS})$,
electron-phonon coupling constant $\lambda$ and critical temperature
$T_c$.}
\label{tableMEP}
\end{table}

It might be possible that MEP could also be computed for other more
complicated superconducting compounds such as heavy fermion materials,
high-$T_c$ cuprates, MgB$_2$, fullerenes etc., with the purpose of having
an indicator  to distinguish them in different categories.
Specially interesting could be the dependence of MEP with $T$ as
we approach the transition temperature $T_c$ from the SC phase.

\noindent {\em Acknowledgments}. This work is partially supported by the
DGES under contract BFM2000-1320-C02-01.

\end{document}